# A MODERN ETHERNET DATA ACQUISITION ARCHITECTURE FOR FERMILAB BEAM INSTRUMENTATION[*]

R. Santucci[†], J. Diamond, N. Eddy, A. Semenov, D. Voy, Fermilab, Batavia, IL, USA

## Abstract

The Fermilab Accelerator Division, Instrumentation Department is adopting an open-source framework to replace our embedded VME-based data acquisition systems. Utilizing an iterative methodology, we first moved to embedded Linux, removing the need for VxWorks. Next, we adopted Ethernet on each data acquisition module eliminating the need for the VME backplane in addition to communicating with a rack mount server. Development of DDCP (Distributed Data Communications Protocol), allowed for an abstraction between the firmware and software layers. Each data acquisition module was adapted to read out using 1 GbE and aggregated at a switch which up linked to a 10 GbE network. Current development includes scaling the system to aggregate more modules, to increase bandwidth to support multiple systems and to adopt MicroTCA as a crate technology. The architecture was utilized on various beamlines around the Fermilab complex including PIP2IT, FAST/IOTA and the Muon Delivery Ring. In summary, we were able to develop a data acquisition framework which incrementally replaces VxWorks & VME hardware as well as increases our total bandwidth to 10 Gbit/s using off the shelf Ethernet technology.

## INTRODUCTION

The current framework of the Fermilab complex, regarding data acquisition, consists of three key areas: hardware platforms, embedded systems, and software. Of the large number of systems, many consist of a VME based platform with a variety of different VxWorks based crate controllers [1]. This architecture has proven extremely reliable; however, it has seen a technical debt build up. Aging crate hardware and stale software has led to rising maintenance costs as new requirements are needed to meet intensity goals and as new projects come online. To meet these new goals, each challenge was worked on iteratively, addressing each bottleneck in the full hardware/software life cycle. The primary hardware hurdle was removing the need of the VME back plane along with a VxWorks based crate controller. Moving to a Linux rack mount server and adopting a 10 GbE network, allowed us to increase bandwidth and decrease complexity in our systems. This was driven by the adoption of Ethernet on each digitizer. The Distributed Data Communication Protocol (DDCP) was developed to abstract an Ethernet field bus interface for the digitizers. This hardware abstraction layer, utilizing the POSIX socket API, allowed for easy adoption and network communication during development. Alongside aging hardware has been software stagnation. This stagnation, was largely driven by the complexity of individual systems with no common platforms. The move to Linux has allowed for the adoption of many new software technologies many of which utilize open-source communities. By transitioning to off the shelf Ethernet, we have been able to drastically increase our data throughput. Installations at the Fast/IOTA complex, the PIPII Test stand, as well as the current roll out to the Muon delivery ring have seen lower user complexity as a direct cause and effect from higher bandwidth. Future development will see the adoption of MicroTCA, standard socFPGA modules, and a common software suite to allow for easy adoption to more systems.

## HARDWARE PLATFORMS

### VME and VxWorks

Current Instrumentation front-ends utilize many different platforms, the primary being VME with a VxWorks RTOS. These systems have been in use ranging from 5 – 30 years. Although reliable, the cost to maintain these systems has become prohibitive. Crate controllers, (MVME500, MVME2400, MVME2401, MVME2434) have either seen drastic price increases or are no longer in production / available. Digitizer stockpiles are dwindling with no clear path to acquire more. Code bases have stagnated as it is prohibitive to update systems to modern compilers. The learning curve of VxWorks makes it difficult to train new developers on aging systems. In addition, data bandwidth requirements have shifted to need larger and larger amounts of data. Limited to a theoretical backplane speed of 40 MB/s (VMEbus IEEE-1014) adopting Ethernet was a less costly transition to increase bandwidth.

### MicroTCA.4

Work is underway to transition from VME to the MTCA.4 platform for instrumentation [2]. In addition to modern crate standards, this platform provides Ethernet fabric on the backplane which is managed by the MTCA Carrier Hub (MCH) with a 10 Gbit uplink. The MTCA.4 standard also provides for a Rear Transition Module (RTM) which can be used to integrate analog signal conditioning which was typically done in an external crate for the VME systems.

### Custom Digitizer

The hardware used to transition the DAQ from VME to Ethernet is a custom 8 channel 250 MS/s 16 bit digitizer board developed at Fermilab in 2016. A block diagram of the module is shown in Fig. 1. The board was originally targeted for VXS with Ethernet on the backplane. Unfortunately, this never came to fruition as the technology became obsolete and the VXS switches were no longer available. For flexibility, a front-panel Ethernet connection was also included on the design and has been the key to enabling the development of the Ethernet DAQ platform.

---

[†] rsantucc@fnal.gov

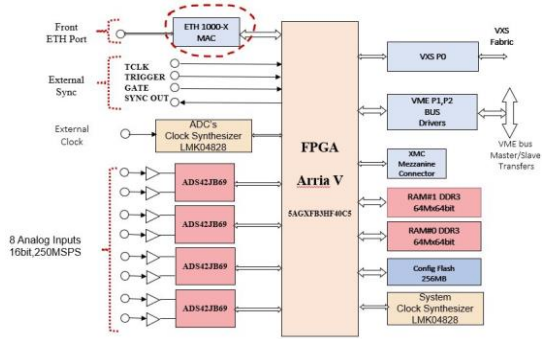

Figure 1: Block Diagram for the Fermilab 8 channel 250 MS/s VME/VXS digitizer. A key feature was the option to provide a Gigabit Ethernet interface from the FPGA to the front panel.

The first implementation of this digitizer was in the IOTA ring BPM system in 2018 [3]. The FAST/IOTA facility is a flexible complex focused on the development of new accelerator technologies. The initial BPM DAQ was a VMEbus system using an x86 slot controller running Linux instead of VxWorks. After initial commissioning, work began on modifying the system to use Ethernet instead of the VMEbus. In this configuration, the digitizers only use the VME crate for power and cooling. The DAQ upgrade occurred in two steps. The first step was to readout over Ethernet by the x86 slot controller. The second step was to replace the slot controller with a standard rack mount server which provided over an order of magnitude resource improvement at a fraction of the cost. The full server and switch configuration was used to readout two crates of digitizers for the PIP-II Injector Test BPM System [4].

The Distributed Data Communication Protocol (DDCP) interface was developed to facilitate the Ethernet interface between the FPGA and the front-end. To simplify the firmware designs, the UDP protocol was selected. The digitizer has an Altera Aria V with a NIOS softcore ARM processor implemented in HDL. The initial DDCP implementation was done in NIOS, but this proved to be a severe bottle neck for large data transfer with data rates <1 MB/s. A dedicated HDL interface was developed to handle high speed data readout. This implementation was able to operate at the full Gigabit Ethernet speed. By utilizing jumbo packets (9 kB) the system achieved real world data rates of 100 MB/s from a single card. This was a significant improvement from the measured rates of 8 MB/s for the same system through VMEbus.

## SOFTWARE

### Linux

From a software perspective, the biggest change has been adopting Linux on our embedded platforms. Moving away from VxWorks to the open-source community has seen easier adoption of new tools, a more gradual learning curve, and the ability to use plug and play off the shelf technologies. Real time applications can be developed using the RTAI framework or by dedicating specific cores to a FreeRTOS kernel.

### DDCP

DDCP was developed as a Ethernet fieldbus protocol [5]. Relying on a request reply structure, data can be transmitted as scalars or structures. All nodes in a given system are given a unique "slot" number. This has been expanded upon to be synonymous to the lower octet of the IP address. Data items, know as features, are accessed and manipulated with various defined operations with data being addressed as slot number, feature number, operation, index, and count. Devices can act as both a client or server on the network which allow for complex structures such as streams and interrupts. The typical architecture is one controller attached to many clients. DDCP can operate in both UDP and TCP mode, with UDP being advantageous when communicating with an FPGA where access to a TCP stack is limited.

### Server

The software architecture can be broken into 2 categories, the server-based software, and the embedded node software. On the server, a Redis database is used to store live and aborted data as well as instrument configuration. Utilizing the different data types within Redis, we built a messaging system (pub/sub), a circular buffer(sets/gets), store configuration (Hash maps) as well as stream live data (Redis streams [6]) as shown in Fig. 2. Communication with embedded nodes is handled via DDCP or Redis TCP. When the systems are triggered, parallel reads can take advantage of the 10 Gbit network and process data asynchronously from the data acquisition once it hits the server. Each process on the server runs within its own container context. These include node monitors, data acquisition modules (bridging Redis and DDCP), multicast listeners, and buffer abort managers to name a few. Data access is handled via a memory API which allows for easy access from control systems (EPICS/ACNET), individual users, as well as system experts for diagnostics. This software suite has allowed for easy adoption of new systems as well as lowered overall monolithic software complexity and has allowed for easier scaling. Future work includes direct streaming into a data lake storage layer as well as building out a dedicated EPICS to Redis driver. In addition, as a benefit to aggregating data in one place, new AI/ML models can be introduced to do server based learning before pushing data to the control system.

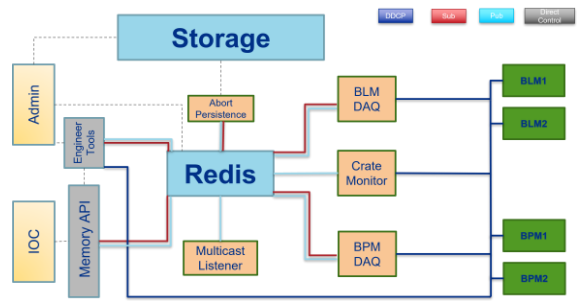

Figure 2: Block Diagram of the Redis system.

Each embedded node has the option of communicating via DDCP, a UDP protocol, or Redis, a TCP protocol. DDCP is utilized in embedded nodes which contain a standalone FPGA which has allowed easier communication between firmware and software. This requires a data acquisition module on the server cluster to bridge to Redis. If a socFPGA is used, the embedded node can communicate directly to Redis utilizing the Linux TCP stack.

## NETWORK ARCHITECTURE

The data acquisition system has been designed to be highly modular. By moving away from the edge model, (data processing at each crate controller), we have been able to lower the complexity of our systems. The acquisition system utilizes either a single rack mount server or can be scaled to include multiple servers. These servers communicate to each system via a 10 Gbit network which is aggregated at house servers. Each house can aggregate any number of digitizers each of which communicates on a 1 Gbit network. Each digitizer acts as an independent client on the network. Each is individually addressable and does not inhibit any other. This model was an easy adoption as we already have Ethernet experience and is extremely cost friendly.

### IOTA, PIP2IT, Muon Delivery Ring, Booster

The first iteration of the system was installed at the Fast/IOTA complex which aggregated 11 custom digitizers, reading out at a total of 2 MB of bursted data. A mirror of the IOTA system was also installed in the PIP2IT test stand. Work is currently being done to install the same system in the Muon Delivery Ring which will be scaled to 30 digitizers and a readout of 350 MB at 1 Hz (Fig. 3). Future work on the Booster BPM systems will see similar installations.

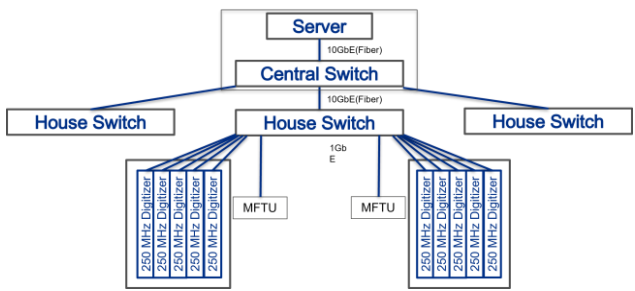

Figure 3: Muon Delivery Ring BPM System which utilizes an MFTU (Multi Function Timing Unit) for triggering and single house switches aggregating 2 VME crates with 5 digitizers at each house.

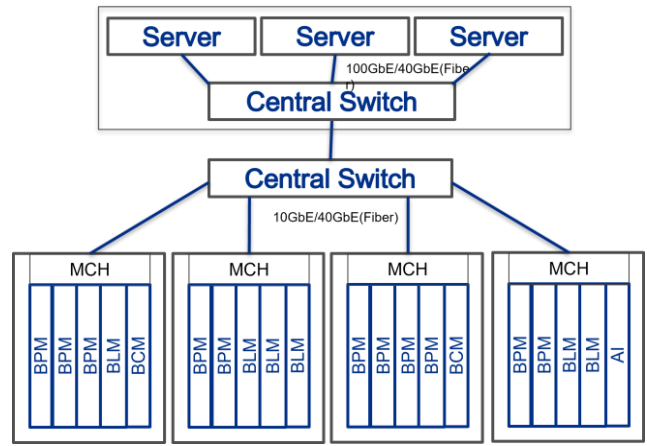

Figure 4: PIP2 Linac which will host multiple instrumentation devices within a crate, each addressable by IP address.

### PIP2

Future work includes standardizing to a common hardware platform for the PIP2 Linac. The instrumentation department has decided to move to MicroTCA with one huge benefit being that it offers Ethernet on the backplane. This allows for the MCH crate handler to act as the 10 Gbit switch aiding in the network architecture (Fig. 4). Another benefit of the system is the ability to hot swap cards, dynamically allocate IP addresses based on slot numbers, as well as host multiple systems in one crate lowering cost. This work is in anticipation of the new PIP2 superconducting Linac in which all instrumentation devices will be front ended through the Ethernet architecture. By expanding to a 100 Gbit network, we anticipate reading out all 1500 instrumentation channels across 200 digitizers in parallel at 20 Hz. This equates to roughly 750 MB of data per second. By clustering multiple servers, load balancing and utilizing hot spares we can drastically improve up time of the system.

### Future Work

With ambitious modernization goals, common platforms and cost saving techniques will be beneficial. By building out a highly modular Ethernet network, older systems can easily be upgraded to utilizing the system. By developing two IP protocols as drop in modules, new instruments and legacy systems can co exists side by side (Fig. 5).

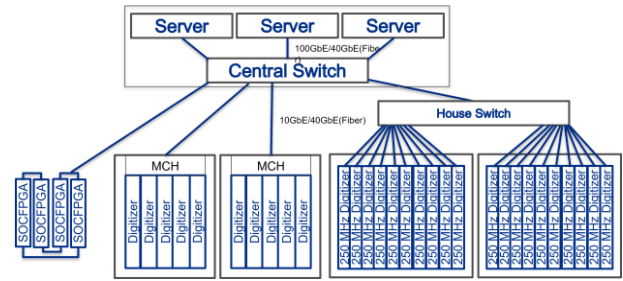

Figure 5: A hypothetical model showing the flexibility of Ethernet as a field bus talking with multiple systems.

## CONCLUSION

The Fermilab Instrumentation department has been successful in its first pass to modernize its data acquisition framework. Systematically updating the complex to utilize off the shelf Ethernet technologies as well as building custom Ethernet modules has been a cost effective way to meet larger and faster data requirements. Initial installations at the IOTA ring and PIP2IT accelerator as well as current installations within the Muon delivery ring have allowed for the development of a software suite which can be used on future systems including the booster bpm system and PIP2 instrumentation system. By moving away from VME and VxWorks, costs and complexity are significantly lowered. This hardware and software standardization will modernize the acquisition system with hopes of easier maintainability and future upgrading.

## ACKNOWLEDGEMENTS

This work was produced by Fermi Research Alliance, LLC under Contract No. DE-AC02-07CH11359 with the U.S. Department of Energy. Publisher acknowledges the U.S. Government license to provide public access under the DOE Public Access Plan